\documentclass[superscriptaddress, reprint, amsmath, amssymb, aps, pre, floatfix]{revtex4-2}
\usepackage{graphicx}
\usepackage{bm}
\usepackage{siunitx}
\usepackage[utf8]{inputenc}
\usepackage[T1]{fontenc}
\usepackage{amsmath,amsfonts,amssymb}
\usepackage{tikz}

\begin{document}
\title{Recovering the activity parameters of an active fluid confined in a sphere}

\author{Cristian Villalobos}
\affiliation{Departamento de Física, FCFM, Universidad de Chile, Santiago, Chile.}
\author{María Luisa Cordero}
\affiliation{Departamento de Física, FCFM, Universidad de Chile, Santiago, Chile.}
\author{Eric Cl\'ement}
\affiliation{Laboratoire PMMH-ESPCI, UMR 7636 CNRS-PSL-Research University, Sorbonne Université, Université Paris Cité, 7-9 quai Saint-Bernard, 75005 Paris, France.}
\affiliation{Institut Universitaire de France, Paris, France.}
\author{Rodrigo Soto}
\affiliation{Departamento de Física, FCFM, Universidad de Chile, Santiago, Chile.}

\begin{abstract}
The properties of an active fluid, for example, a bacterial bath or a collection of microtubules and molecular motors, can be accessed through the dynamics of passive particle probes. Here, in the perspective of analyzing experimental situations of confinement in droplets, we consider the kinematics of a negatively buoyant probe particle in an active fluid, both confined within a spherical domain. The active bath generates a fluctuating flow that pushes the particle with a velocity that is modeled as a colored stochastic noise, characterized by two parameters, the intensity and memory time of the active flow.
When the particle departs a little from the bottom of the spherical domain, the configuration is well approximated by a particle in a two-dimensional harmonic trap subjected to the colored noise, in which case an analytical solution exists, which is the base for quantitative analysis. We numerically simulate the dynamics of the particle and use the planar, two-dimensional mean square displacement to recover the activity parameters of the bath. This approach yields satisfactory results as long as the particle remains relatively confined, that is, as long as the intensity of the colored noise remains low.
\end{abstract}

\date{\today}

\maketitle

\section{Introduction}

When characterizing a fluid flow, the use of tracer particles is a common approach. In particle image velocimetry (PIV) and particle tracking velocimetry (PTV), for example, the flow is seeded with tracer particles whose velocities are measured by optical means, under the assumption that they reflect the velocity of the carrier fluid~\cite{RaffelPIV}. In general, neutrally buoyant small particles follow the flow more accurately. The same principle has been used to characterize bacterial flows. In those cases, the tracer particles are usually of the same size or larger than the bacteria, and thus, although micrometric in size, they can hardly follow the flow streamlines. Instead, particles interact with the hydrodynamic fields produced by the swimming bacteria~\cite{hernandez2005transport, Thiffeault2010, pushkin2013fluid, morozov2014enhanced, Kasyap2014} and undergo random sequences of binary collisions with them~\cite{gregoire2001active, lagarde2020colloidal}. The resulting particle dynamics thus becomes an indirect probe for the statistical properties of the bacterial bath~\cite{chen2007fluctuations, maggi2017memory, Tripathi2022, seyforth2022nonequilibrium}.

A suspension of swimming bacteria, performing frequent active random reorientations, and reorienting also due to the thermal noise and the flow field generated by other bacteria, produces a fluctuating flow in the external liquid medium that can be described as an active noisy bath. One effect of this active bath on a tracer particle is an increased diffusivity~\cite{wuParticleDiffusionQuasiTwoDimensional2000, gregoire2001active, kim2004enhanced, hernandez2005transport, chen2007fluctuations, leptos2009dynamics, Thiffeault2010, mino2011enhanced, zaid2011levy, jepson2013enhanced, pushkin2013fluid, Kasyap2014, morozov2014enhanced, patteson2016particle, peng2016diffusion, maggi2017memory, lagarde2020colloidal, xie2022activity}, which has been described as an increased effective temperature~\cite{wuParticleDiffusionQuasiTwoDimensional2000, zaid2011levy, parra-rojasActiveTemperatureVelocity2013, szamelSelfpropelledParticleExternal2014, patteson2016particle, maggi2017memory, dabelow2019irreversibility, lagarde2020colloidal}. Although the concept of an effective temperature is easy to grasp, the noisy flow caused by the active bath is not thermal but instead can be better characterized as an exponentially correlated colored noise with at least \emph{two} independent activity parameters, that here we choose to define as the noise intensity and its persistence time at a fixed position~\cite{argunNonBoltzmannStationaryDistributions2016, caprini2018active, chaki2018entropy, chaki2019effects, dabelow2019irreversibility, goswami2019heat, goswami2019diffusion, yeActiveNoiseExperienced2020}. By immersing a probe particle in the bacterial suspension, the noisy bath exerts a fluctuating drag force on the passive particle that yields a fluctuating movement, from which the bath parameters could be, in principle, obtained.

One simple conceptual configuration that serves this objective is a passive particle immersed in the bacterial fluid and confined in a harmonic potential, such as an optical trap~\cite{chen2007fluctuations, maggiGeneralizedEnergyEquipartition2014, argunNonBoltzmannStationaryDistributions2016, chaki2018entropy, chaki2019effects, goswami2019heat, goswami2019diffusion, das2018confined, yeActiveNoiseExperienced2020, seyforth2022nonequilibrium, paul2022force}. Here, the simple form of the forces acting on the particle due to the harmonic potential and the active flow, which is modeled as a colored noise, makes it possible to find an analytical solution for the mean square displacement (MSD) of the tracer~\cite{maggiGeneralizedEnergyEquipartition2014}. Then, the statistical properties of the bath can be obtained by fitting the dynamics of the particle, to the model prediction for the MSD.

A similar experimental setup consists in confining the active suspension inside a circular or spherical domain together with a passive probe particle. This configuration was theoretically considered~\cite{Freund2023} and experimentally implemented in a circular capillary~\cite{maggiGeneralizedEnergyEquipartition2014, maggi2017memory} and in a double emulsion droplet~\cite{yodh2023zonal}. Other geometries for the confinement have been considered as well~\cite{ye2024particle}. If the probe particle is not neutrally buoyant, then sedimentation due to gravity will compete with the fluctuating active forcing. As a result, the probe particle will explore the spherical domain around the equilibrium position that it would display in absence of the active fluid (hereafter the ``equilibrium position'' for simplicity). From the statistics of this dynamics, the properties of the active bath can be obtained, just like the optical trap case.

In this work, we consider the above-mentioned scenario of a negatively buoyant passive tracer particle immersed in an active fluid under spherical confinement. The objective is to answer the question: how can we formally extract the two  parameters characterizing bacteria activity from the dynamics of the tracer particle? For that, through numerical simulations, we fix the activity parameters of the active fluid and record the motion of a passive particle. Then we try to recover the activity parameters from a fit of the particle in-plane MSD with a curve drawn from the model that describes a particle in a two-dimensional harmonic potential subjected to an active colored noise. Due to sedimentation, the spherical confinement is well approximated by a harmonic potential only when the particle departs little from the equilibrium position and in that condition, the parameters characterizing the activity are directly recoverable. However, care must be taken to adequately capture the properties of the bath and, in particular, to distinguish between the persistence time of the active bath and the characteristic sedimentation timescale. Beyond that limit, we also show that the activity parameters extracted from this model can reasonably describe the active bath properties, at least in some regions of the parameter space.

\section{Three-dimensional simulations} \label{sec.Simulations}

\subsection{Model}

\begin{figure}[ht]
\includegraphics[width=6.5cm]{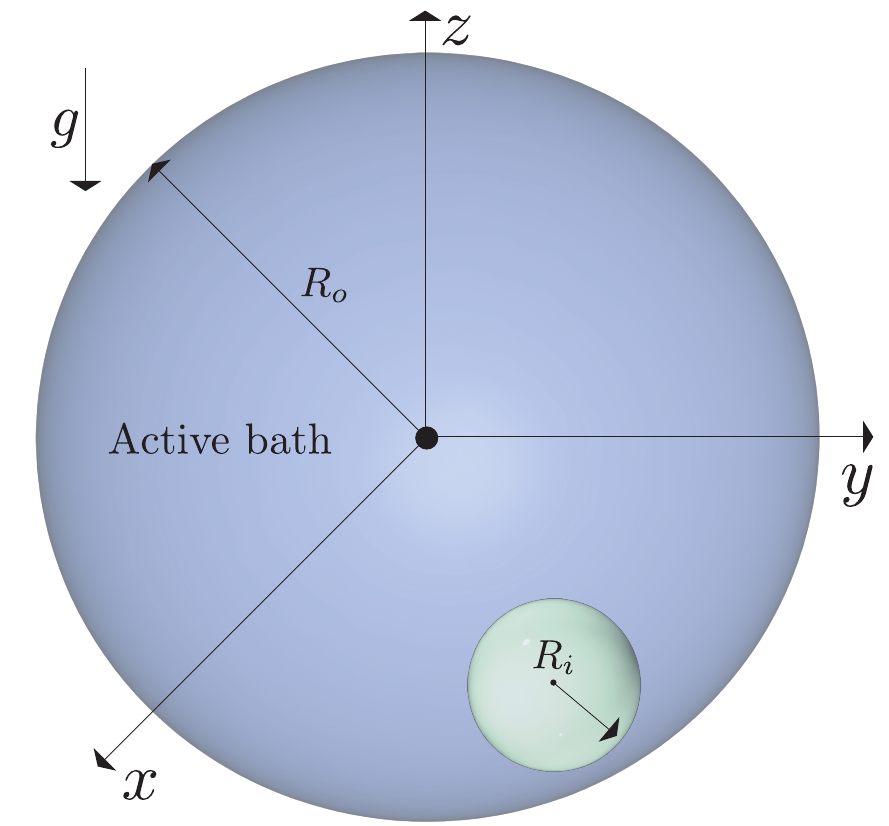}
\caption{Sketch of the system. A passive particle of radius $R_i$ inside a sphere of radius $R_o$ filled with an active bath. The passive particle is subject to gravity pointing toward negative $z$.}
\label{fig.sim_exp}
\end{figure}

We consider a non-Brownian passive spherical particle of radius $R_i$ suspended in an active fluid, with both the particle and the active fluid confined in a spherical domain of radius $R_o$ (see Fig.~\ref{fig.sim_exp}). The activity of the fluid generates a fluctuating velocity field that drives the motion of the particle. For small $R_i$, in the point tracer limit, the induced velocity of the particle is simply the flow velocity at the particle position. For slightly larger particle sizes, the Fax\'en correction should be included~\cite{rallison1978note, happel1983low}. For even larger particles, the particle velocity results from solving the flow field in presence of the active stresses generated by the bacteria. To close the problem, the boundary conditions at the particle surface and at the border of the enclosing sphere should be explicitly considered. In any of these cases, the active suspension imposes a fluctuating velocity $\mathbf{u}(t)$ on the particle. On top of this, due to the gravitational acceleration $g$, the particle sediments with a speed $v_s = \Delta M g / \gamma$, where $\Delta M$ is the buoyant mass and $\gamma$ is the friction coefficient. Depending on the sign of $\Delta M$, the particle can sediment along or against gravity. For simplicity and without loss of generality, we will consider  that $\Delta M$ is positive, implying that the equilibrium position is at the bottom of the sphere. As a result of the sedimentation and the fluctuating field, the equation of motion for the particle with position $\mathbf{r} = (x, y, z)$ is
\begin{equation}
\label{eq.Numeric_model_Dim}
\mathbf{\dot r} = \mathbf u(t) - v_s \mathbf{\hat e}_z,
\end{equation}
with boundary condition $x^2 + y^2 + z^2 \leq (R_o - R_i)^2$, where we have assumed that the particle moves in the low Reynolds number regime~\cite{happel1983low}.

For the active bath, we model $\mathbf{u}$ as a Gaussian noise with amplitude $u_b$ and persistence time $\tau_b$, $\langle u_i(t) u_j(t') \rangle = u_b^2 \delta_{ij} e^{-|t - t'| / \tau_b}$, with $i, j = x, y, z$ . That is, we model it as an Ornstein--Uhlenbeck process (OUP) following the dynamical equation:
\begin{equation}
\dot{\mathbf u} = - \frac{\mathbf u}{\tau_b} + \boldsymbol \eta(t),
\end{equation}
where $\boldsymbol \eta$ is a white noise with zero mean and delta correlated in time, $\langle \eta_i(t) \eta_j(t') \rangle = 2 u_b^2 / \tau_b \delta_{ij} \delta(t - t')$. This model captures the essential features of the bacterial swimming mechanism, including persistence in motion and random reorientations~\cite{argunNonBoltzmannStationaryDistributions2016, caprini2018active, chaki2018entropy, chaki2019effects, dabelow2019irreversibility, goswami2019heat, goswami2019diffusion, yeActiveNoiseExperienced2020}.

To analyze the resulting Langevin equation with colored noise, we define the dimensionless position $\tilde{\mathbf{r}} = \mathbf{r} / R$ and time $\tilde{t} = t v_s / R$, where $R = R_o - R_i$. Here, we opted to make time dimensionless with $\tau_s = R/v_s$, the characteristic sedimentation time, because it is a time scale that can be accessed experimentally, as it also happens for the characteristic length scale $R$. With this, we define the two dimensionless control parameters: the dimensionless bath persistence time
\begin{align}
\tilde \tau_b = \tau_b / \tau_s
\end{align}
and bath persistence length
\begin{align}
\tilde \ell_b = u_b \tau_b / R.
\end{align}
In terms of them, the dimensionless bath speed is $\tilde u_b = u_b / v_s = \tilde \ell_b / \tilde \tau_b$.

In summary, the system is described by two dimensional parameters, $R$ and $v_s$ that determine the degree of confinement and the speed of sedimentation, and two dimensionless parameters, $\tilde \ell_b$ and $\tilde \tau_b$ that compare the persistence length and time of the active fluid to the particle's confinement and sedimentation time. The resulting dimensionless equation of motion for the particle is (dimensionless variables are denoted by tildes):
\begin{equation}
\label{eq.Motion_dimless}
\frac{d \mathbf{\tilde r}}{d \tilde t} = \mathbf{\tilde u} - \hat{\mathbf e}_z,
\end{equation}
where now the boundary condition for the particle reads:
\begin{equation}
\label{eq.BC}
|\mathbf{\tilde r}|^2 = \tilde x^2 + \tilde y^2 + \tilde z ^2 \leq 1.
\end{equation}
The imposed velocity satisfies the Langevin equation for the OUP
\begin{equation}
\label{eq.AOUP}
\frac{d \mathbf{\tilde u}}{d \tilde t} = - \frac{\mathbf{\tilde u}}{\tilde \tau_b} + \boldsymbol{\tilde \eta} \left( \tilde t \right),
\end{equation}
where the correlation of the non-dimensional noise is $\langle \tilde \eta_i (\tilde t) \tilde \eta_j( \tilde t') \rangle  = 2 \tilde \ell_b^2 \tilde \tau_b^{-3} \delta_{ij} \delta(\tilde t - \tilde t' ) = 2 \tilde u_b^2 \tilde \tau_b^{-1} \delta_{ij} \delta(\tilde t - \tilde t' )$.

The memoryless case, that is, the tracer subject to Brownian noise, corresponds to taking the limit $\tilde{\tau}_b \to 0$ and $\tilde{\ell}_b \to 0$, while keeping $\tilde{\ell}_b^2/\tilde{\tau}_b$ fixed, equal to the diffusion coefficient.

\subsection{Simulation results}

\begin{figure*}[hb]
\begin{tikzpicture}

 \node at (0, 0) {\includegraphics[width=.32\textwidth]{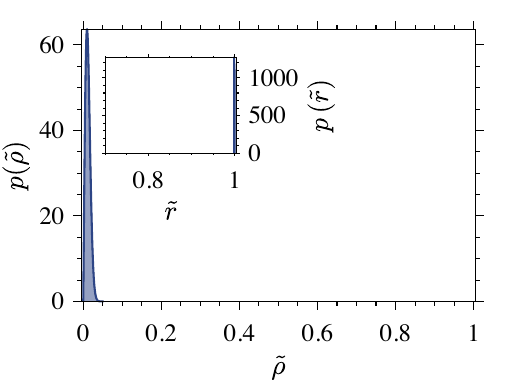}};
 \node at (6, 0) {\includegraphics[width=.32\textwidth]{
  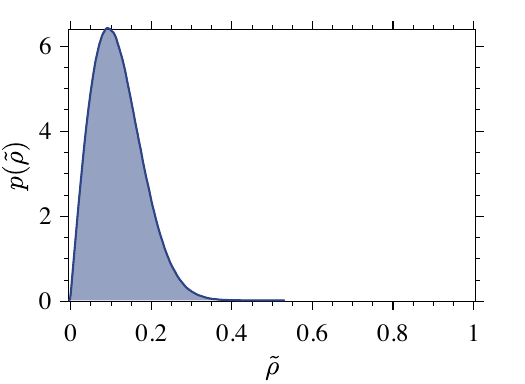}};
 \node at (12, 0) {\includegraphics[width=.32\textwidth]{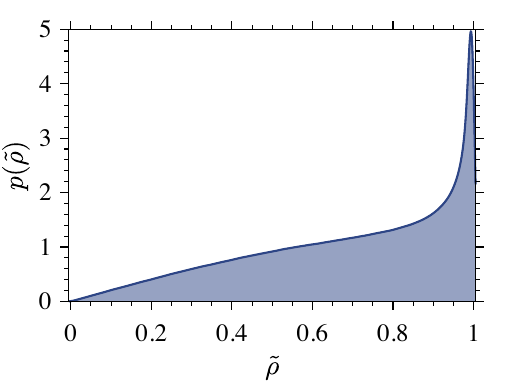}};

 \node at (0, -10) {\includegraphics[width=.32\textwidth]{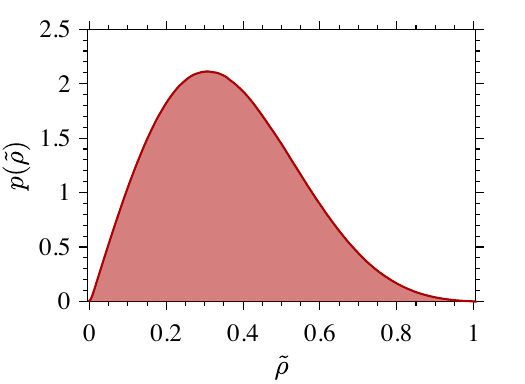}};
 \node at (6, -10) {\includegraphics[width=.32\textwidth]{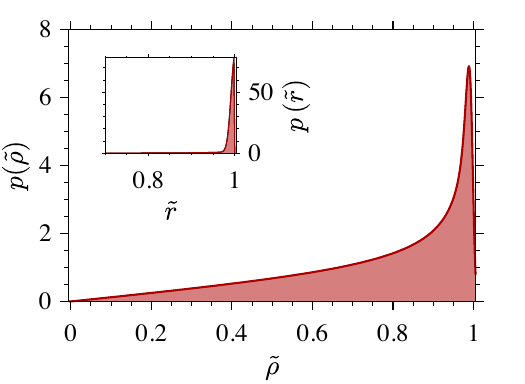}};
 \node at (12, -10) {\includegraphics[width=.32\textwidth]{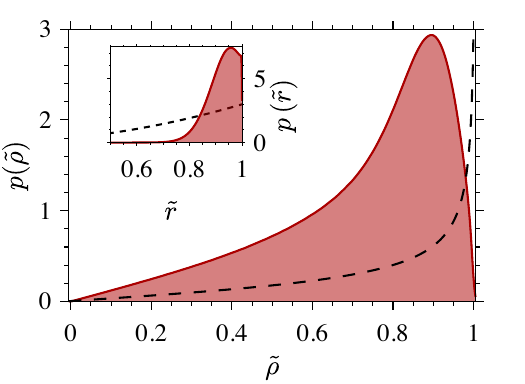}};

 \node at (0, -5) {\includegraphics[width=.32\textwidth]{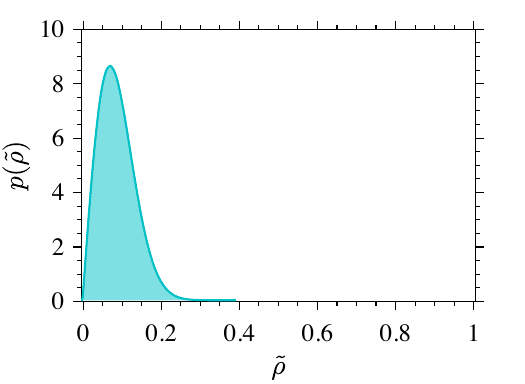}};
 \node at (6, -5) {\includegraphics[width=.32\textwidth]{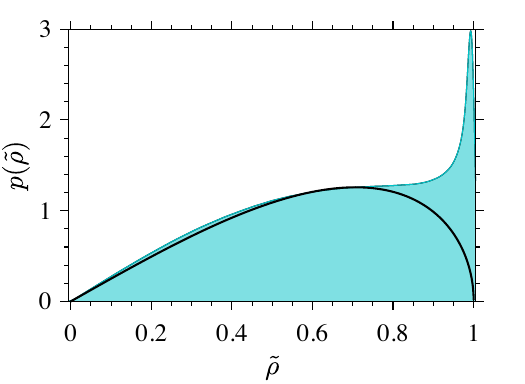}};
 \node at (12, -5) {\includegraphics[width=.32\textwidth]{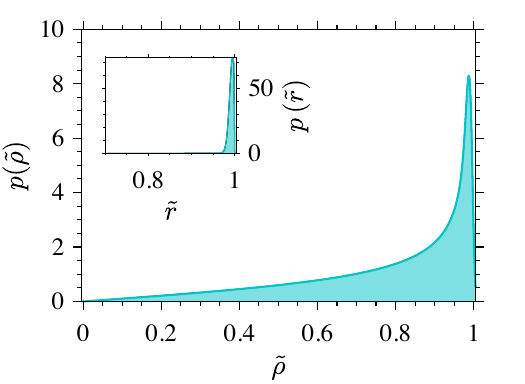}};

 \draw[->, thick] (-3, -12.5) -- (14.5, -12.5) node[midway, fill=white] {$\tilde{\ell}_b$};
 \draw[->, thick] (-3, -12) -- (-3, 2) node[midway,fill=white] {$\tilde{\tau}_b$};

\node at (-2.1,2.1) {(a)};
\node at (3.7,2.1) {(b)};
\node at (9.6,2.1) {(c)};

\node at (-2.1,-2.9) {(d)};
\node at (3.7,-2.9) {(e)};
\node at (9.6,-2.9) {(f)};

\node at (-2.1,-7.9) {(g)};
\node at (3.7,-7.9) {(h)};
\node at (9.6,-7.9) {(i)};
\end{tikzpicture}

\caption{Probability density function for the planar dimensionless position, $p(\tilde \rho)$ for different values of the control parameters. From bottom to top: $\tilde \tau_b = 0.1$ (red), $1.0$ (cyan), and 10 (blue), and from left to right: $\tilde \ell_b = 0.1$, 1.0, and 10. Insets: Probability density function of the three-dimensional dimensionless radius, $p(\tilde r)$. For the cases where the inset is not shown, $p(\tilde r)$ is essentially equal to the (a) subplot. The solid and dashed black lines in (e) and (i) represent the theoretical probability distributions of the planar radius for a particle uniformly distributed on the inner surface and in the volume of a sphere [$p_\text{unif.surf.}(\tilde \rho)$ and $p_\text{unif.vol.} (\tilde \rho)$], respectively. The dashed line in the inset of (i) shows the theoretical probability distribution of the three-dimensional dimensionless radius for a particle uniformly distributed in the volume of a sphere, $p_\text{unif.vol.}(\tilde r)$.}
\label{fig.PDF}
\end{figure*}
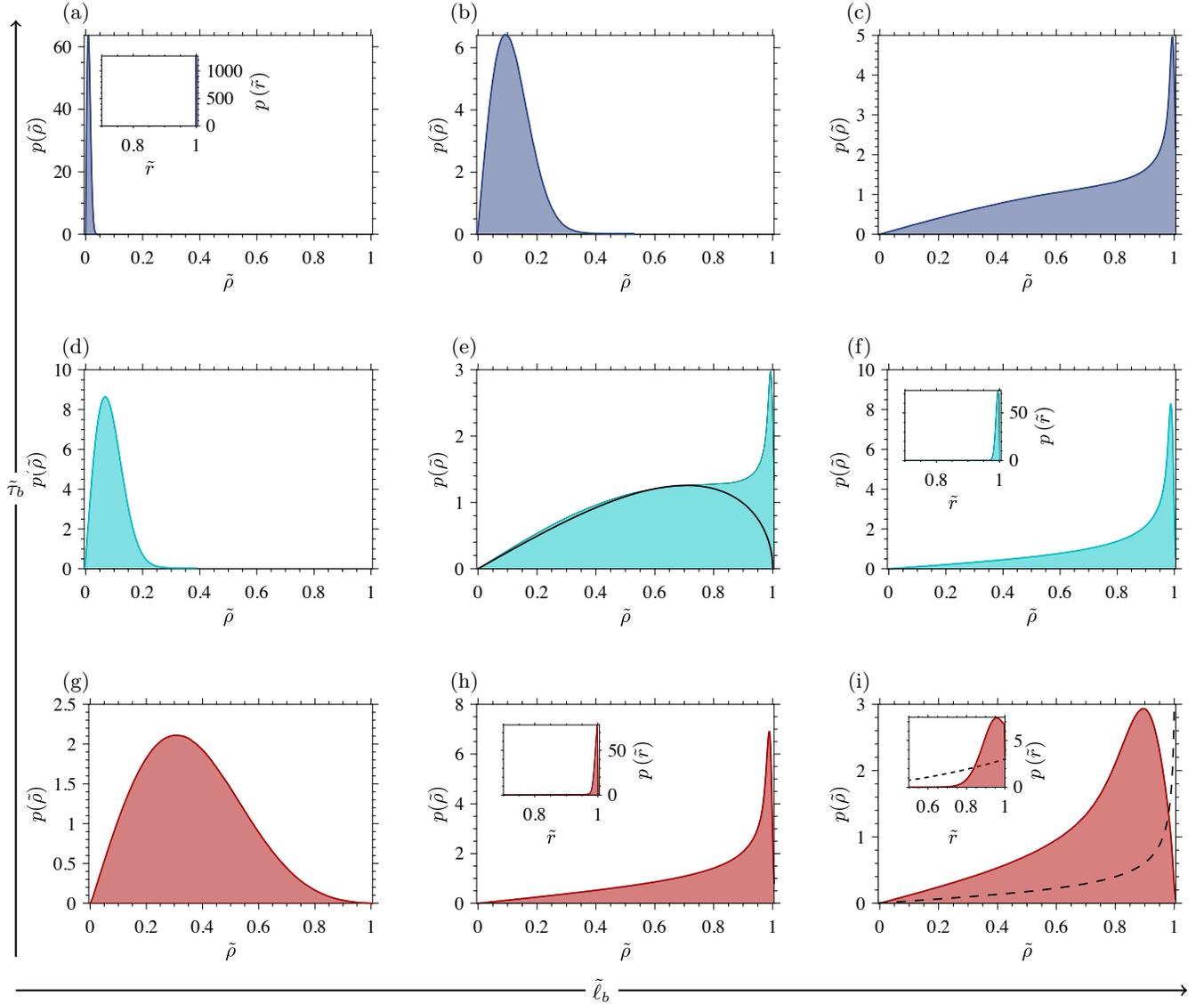

In order to investigate the system, we numerically solve the dynamical equations~\eqref{eq.Motion_dimless} and \eqref{eq.AOUP} with a forward Euler scheme. To confine the particle [Eq.~\eqref{eq.BC}], we use the reflective boundary condition method~\cite{Volpe2014}. Simulations are ran for $\tilde \tau_b$ and $\tilde \ell_b$ in the range $[0.1-10]$, which is evenly explored in a logarithmic scale. The particle is placed initially at the equilibrium position, $\tilde{\mathbf{r}}_\text{eq} = - \hat{\mathbf{e}}_z$. For each set of parameters, we first determine empirically the time needed for the MSD to saturate (see below). Then, the system is let to relax for this time and, after that, measurements are made for a period at least ten times larger. In the Supplementary Material we present, for illustration, sample trajectories of the particle for relevant sets of parameters.

\begin{figure*}[ht]

\begin{tikzpicture}

\node at (0, 0){\includegraphics[width=0.32\textwidth]{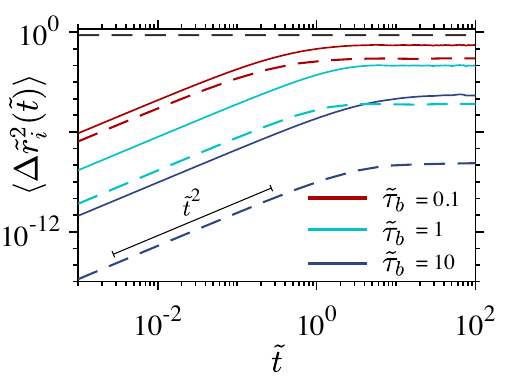}};
\node at (6, 0){\includegraphics[width=0.32\textwidth]{
  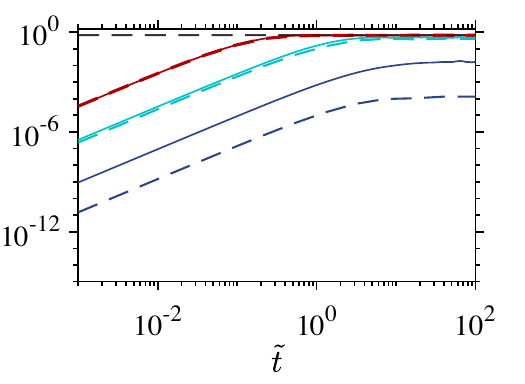}};
\node at (12, 0){\includegraphics[width=0.32\textwidth]{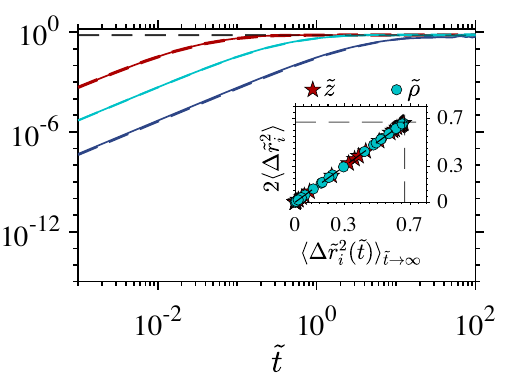}};

\node at (0, -4.5){\includegraphics[width=0.32\textwidth]{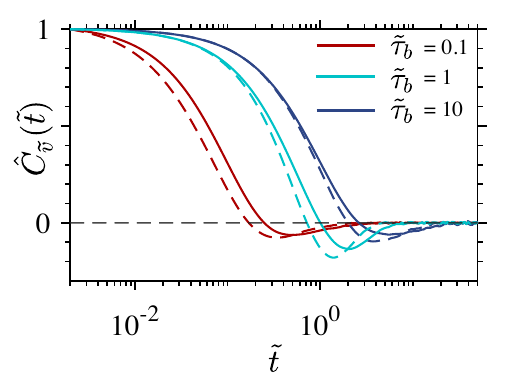}};
\node at (6, -4.5){\includegraphics[width=0.32\textwidth]{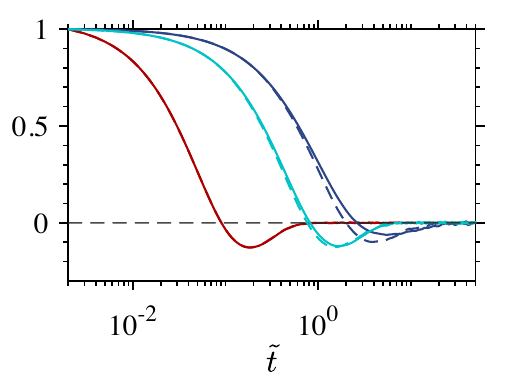}};
\node at (12, -4.5){\includegraphics[width=0.32\textwidth]{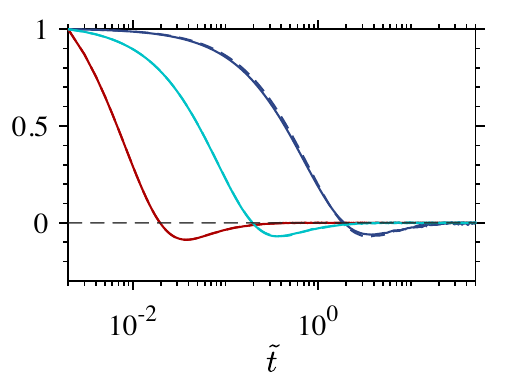}};

\draw[->, thick] (-3, -7) -- (14.5, -7) node[midway, fill=white] {$\tilde{\ell}_b$};

\node at (-2.1,2.1) {(a)};
\node at (3.7,2.1) {(b)};
\node at (9.6,2.1) {(c)};

\node at (-2.1,-2.4) {(d)};
\node at (3.7,-2.4) {(e)};
\node at (9.6,-2.4) {(f)};
\end{tikzpicture}

\caption{(a)--(c): Mean square displacement. (d)--(f): Time-delayed velocity correlation function. The solid lines are the simulation results for the average in the $\tilde x$ and $\tilde y$ coordinates, while the dashed lines are for the $\tilde z$ coordinate. The horizontal black dashed lines in the top panels are placed at the maximum possible saturation MSD value, equal to 2/3. In the bottom panels, the black dashed lines show $C_{\tilde v} = 0$ for reference. From left to right $\tilde \ell_b = 0.1, 1.0, 10$ and, in each panel, $\tilde \tau_b = 0.1$ (red), 1.0 (cyan), and 10 (blue). The inset in (c) shows the linear relation between the variance and the saturation of the MSD for the nine combinations of $\tilde \ell_b$ and $\tilde \tau_b$ presented in the plots.}
\label{fig.MSD_VACF_PACF}
\end{figure*}

Depending on the values of the active fluid parameters, the particle position departs a little from the equilibrium position (see supplementary video S1), or on the other end explores the spherical confinement, whether almost all the time in contact with the inner surface of the sphere (supplementary video S2) or leaving it to explore fully the bulk of the active fluid (supplementary video S3). To analyze the different regimes, we present the probability density functions (PDF) of the particle's position as a function of $\tilde \tau_b$ and $\tilde \ell_b$ in Fig.~\ref{fig.PDF}. Experimentally, for example using a microscope to observe the tracer dynamics, normally only the planar $x$ and $y$ coordinates are accessible with precision. Then, it becomes relevant for practical reasons to obtain the statistical properties of the planar radius $\tilde \rho = \sqrt{{\tilde x}^2 + {\tilde y}^2}$. Figure~\ref{fig.PDF} presents the probability density function $p(\tilde \rho)$ for relevant values of the control parameters. It is seen that for some values of the parameters, the distribution is  peaked at $\tilde \rho \approx 0$, meaning that the particle hardly departs from the equilibrium position, while for other cases a peak appears at $\tilde \rho \approx 1$, which indicates that the particle is exploring the whole spherical space. However, this distribution alone does not indicate whether the particle moves in contact with the inner surface or if it departs from it. To discriminate between these possibilities, we compute $p(\tilde r)$, the probability density function of the distance from the center $\tilde r = |\mathbf{\tilde r}|$, which is shown as insets in Fig.~\ref{fig.PDF}. For the subplots where the inset is not shown, the probability distribution function is essentially equal to the (a) subplot, with $p(\tilde r)$ strongly peaked at $\tilde r \approx 1$, meaning that the particle is almost all the time in contact with the inner surface. To help the analysis, we also present for comparison the theoretical distributions for the planar radius $p_\text{unif.vol.}(\tilde \rho) = 3 \tilde \rho \sqrt{1 - \tilde \rho^2}$ and $p_\text{unif.surf.}(\tilde \rho) = \tilde \rho / \sqrt{1 - \tilde \rho^2}$, for a particle that is uniformly distributed inside the sphere or uniformly distributed on the inner surface, respectively, and the distribution for the distance to the center $p_\text{unif.vol.}(\tilde r) = 3 \tilde r^2$, for a particle that is uniformly distributed inside the sphere.

It is found that as $\tilde \ell_b$ increases, the particle is more likely to be found far from $\tilde \rho = 0$, as an increasing persistence length makes it easier for the bath to take the particle away from the equilibrium position. On the other hand, for values of $\tilde \ell_b \lesssim \tilde \tau_b$ (that is, for $\tilde u_b \lesssim 1$), the particle mainly fluctuates around the equilibrium position as the bath push is weaker than the tendency to sediment (Figs.~\ref{fig.PDF}a, b, d). Almost in all conditions, the particle moves very close to the surface. The shape of $p(\tilde \rho)$ becomes more distributed in $\tilde \rho$ when $\tilde \ell_b \gtrsim \tilde \tau_b$ but $p(\tilde r)$ still remains highly peaked at $\tilde r = 1$ (Figs.~\ref{fig.PDF}e--h), suggesting that in those cases the particles explores the confinement but being almost all the time in contact with the inner surface of the sphere. On the contrary, the particle explores the interior of the sphere for large $\tilde \ell_b$ and small $\tilde \tau_b$, that is, for large $\tilde u_b$, although for the larger values of $\tilde u_b$ explored here, it is still more likely for the particle to be located near the surface (Fig.~\ref{fig.PDF}i).

To study the dynamics of the particle, we also compute the MSDs for each direction, defined as $\left \langle \Delta \tilde r_i^2 (\tilde t) \right \rangle = \left \langle \left[ \tilde r_i (\tilde t + \tilde t_0) - \tilde r_i (\tilde t_0) \right]^2 \right \rangle$, averaged over $\tilde t_0$, with $i = \tilde x, \tilde y, \tilde z$. By symmetry, the $\tilde x$ and $\tilde y$ components are equivalent and, therefore, we present their average. The results are shown in Fig.~\ref{fig.MSD_VACF_PACF}a--c. Two clear regimes can be observed for all the values of the control parameters: ballistic at short times and saturation at long times due to the confinement. The crossover time between these two regimes shows dependence with both parameters, increasing with $\tilde \tau_b$ and decreasing with $\tilde \ell_b$. Indeed, if the bath shows larger persistence times, the particle is also pushed for longer times in the same direction, increasing the duration of the ballistic regime. For the dependence with $\tilde \ell_b$, we recall that increasing this parameter while keeping $\tilde \tau_b$ fixed is equivalent to increasing the bath speed $\tilde u_b$, in which case it takes less time for the particle to explore the sphere and then, the saturation regime of the MSD is  more rapidly achieved. Diffusion, meaning a linear increase of the MSD with time over an appreciable lapse of time, is only observed for $\tilde \tau_b$ smaller than $10^{-3}$ (not shown), which corresponds then to an effective Brownian particle. From the ballistic regime, we can extract the mean square velocity of the particle $\langle \tilde v_i^2 \rangle$ (offset of the log-log plot of the MSD in the ballistic regime), which increases with $\tilde \ell_b$ and decreases with $\tilde \tau_b$. That is, $\langle \tilde v_i^2 \rangle$ increases with the bath velocity $\tilde u_b$. For small values of $\tilde \ell_b$, when the particle barely departs from the equilibrium position (see Fig.~\ref{fig.PDF}), the saturation value at long times decreases with $\tilde \tau_b$, again reflecting its natural increase with the bath velocity $\tilde u_b$. On the other hand, when $\tilde \ell_b$ is large, the saturation is independent of $\tilde \tau_b$, taking the maximum value (see below), reflecting that gravity is irrelevant. Also, for $\tilde \ell_b = 10$, the MSD on $\tilde z$ is almost identical to the MSD on $\tilde x$ and $\tilde y$, meaning that the particle is moving equally in all directions, as a consequence of the irrelevance of gravity. Finally, note that for a confined system, the MSD saturation value in any coordinate equals twice the variance of the position on that coordinate. Indeed, $\lim_{\tilde t \to \infty} \langle \Delta \tilde r_i^2({t})\rangle = \lim_{\tilde t \to \infty} \left[\langle \tilde r_i^2(\tilde t + \tilde t_0)\rangle + \langle \tilde r_i^2({\tilde t}_0)\rangle - 2 \langle \tilde r_i({\tilde t}+{\tilde t}_0) \tilde r_i({\tilde t})\rangle  \right] = 2 \langle \tilde r_i^2 \rangle$, where we used that for long times the positions are uncorrelated. This relation is shown in the inset of Fig.~\ref{fig.MSD_VACF_PACF}c. For a particle confined in a sphere, the position variance is naturally bounded. Assuming that the particle distribution is isotropic, the maximum value for the variance is when the particle uniformly moves on the \emph{surface} of the sphere, in which case the variance on each coordinate is $R^2/3$. Consequently, in Fig.~\ref{fig.MSD_VACF_PACF}, the dimensionless MSDs are compared with 2/3, which is their maximum possible value.

According to the Green--Kubo relation, the long-term diffusion coefficient of the particle is given by the integral of its time delayed velocity correlation function, $C_{\tilde vi}(\tilde t) = \langle {\tilde v}_i (\tilde t_0) {\tilde v}_i (\tilde t + \tilde t_0) \rangle$. As the spherical confinement imposes a null diffusion coefficient for long times, as evidenced in the saturation of the MSD for long times, the integral of $C_{\tilde vi} (\tilde t)$ must vanish identically. Given that $C_{\tilde vi}(0) = \langle \tilde v_i^2\rangle$ is strictly positive, the velocity correlation function must necessarily take negative values during some periods of time. Figure~\ref{fig.MSD_VACF_PACF}d--f presents the results of the simulations for different values of the parameters. In all cases, $C_{\tilde vi} (\tilde t)$ decreases to zero non-monotonically, passing through a single negative minimum. Consistent with the MSD, the dynamics in $\tilde z$ is equal to that of $\tilde x$ and $\tilde y$ when $\tilde \ell_b$ is large.

\section{Harmonic approximation}

\begin{figure}
\includegraphics[width=\linewidth]{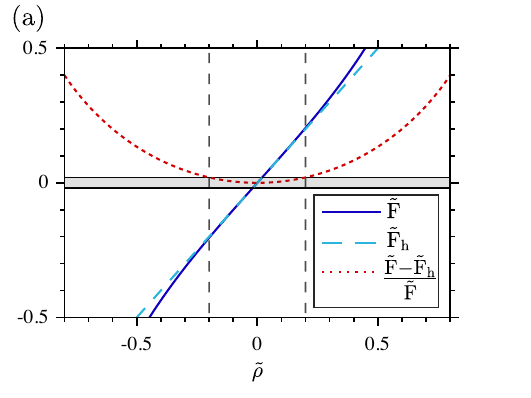}
\includegraphics[width=\linewidth]{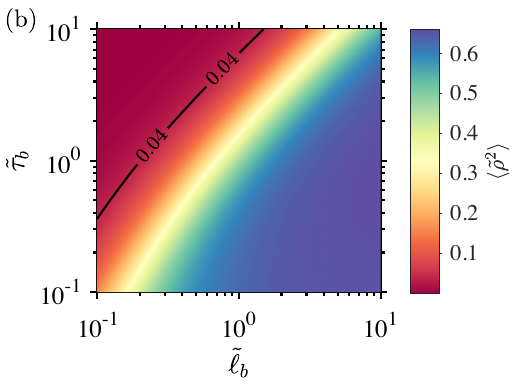}

\caption{Top: Comparison of the normalized restoring force (nondimensionalized with $\Delta M g/R$) as a function of the planar radius $\tilde \rho$ for the particle moving on the surface of the sphere using the full expression (solid blue line) and the harmonic approximation (dashed cyan line) for the gravitational potential. The relative error is shown with a red dotted line, where the threshold of 2\% is indicated with a horizontal gray band. The associated threshold in the planar radius is shown with vertical lines at $\tilde \rho \approx 0.20$. Bottom: Variance of $\tilde \rho$ as a function of the control parameters $\tilde \ell_b$ and $\tilde \tau_b$. As a reference, the solid line indicates the threshold $\langle \tilde \rho^2 \rangle = 0.20^2 = 0.04$ above which the relative error in the force is smaller than 2\% on average. }
\label{fig.variance-aprox}
\end{figure}

When the active bath is weak enough compared to gravity such as to keep the particle near the equilibrium position, the spherical confinement can be approximated by a harmonic trap. Indeed, for a particle that moves in contact with the surface, the gravitational potential $U = \Delta Mg z$ becomes a function only of the planar radius $\rho$ because we can substitute $z = \pm \sqrt{R^2 - \rho^2}$. For small values of $\rho$, assuming a downward sedimentation of the particle, it is possible to expand this expression hence resulting in a harmonic potential $U \approx - \Delta MgR + \Delta Mg \rho^2 / (2R)$. With this approximation, an effective linear restoring force $\mathbf{F}_{\text{eff}} = - k \boldsymbol{\rho}$ results, with $k = \Delta Mg/R$, the harmonic strength. Figure~\ref{fig.variance-aprox}a shows the quality of this approximation, where it is shown that the relative error in the resulting force on the particle is smaller than 2\% when $\rho \lesssim 0.20 R$. In Fig.~\ref{fig.variance-aprox}b, we present the variance of $\tilde \rho$ in the parameter space $(\tilde \ell_b, \tilde \tau_b)$, obtained from the numerical simulations of Sec.~\ref{sec.Simulations}, indicating as a reference the threshold $\langle \tilde \rho ^2 \rangle = 0.20^2 = 0.04$ found above. The variance increases with $\tilde \ell_b$ and decreases with $\tilde \tau_b$. Thus, the harmonic approximation seems appropriate for a wide range of parameters for $\tilde \tau_b$ large and $\tilde \ell_b$ small, that is, when $\tilde u_b$ is small.

Therefore, in this limit, this results suggests that it is legitimate to consider that the motion of the confined particle can be modeled by an effective two-dimensional dynamics of a particle experiencing a spring-like force in the $(x, y)$ plane. The advantage of this model, compared with the full three-dimensional confined system, is that it can be solved analytically. Moreover, it has been used in experimental and numerical configurations to describe particles confined by optical traps under the influence of an active bath \cite{maggiGeneralizedEnergyEquipartition2014, yeActiveNoiseExperienced2020, argunNonBoltzmannStationaryDistributions2016}. Again, $R$ is used as a characteristic length and $\tau_s$ as a characteristic time to nondimensionalize the variables. The equation of motion for the particle in the harmonic approximation is:
\begin{align}
\label{eq.spring_model}
\dot{\tilde \rho}_i &= \tilde u_i (\tilde t) - \frac{\tilde \rho_i}{\tilde \tau_\text{sh}}, &  i &= \tilde x, \tilde y
\end{align}
where we have defined $\tau_\text{sh} = \gamma/k$, the relaxation time associated with the confinement in the harmonic model. Although formally the expression is the same as for $\tau_s$, we keep the alternative notation to differentiate between the models. The dimensionless $\tilde \tau_\text{sh} = \tau_\text{sh}/\tau_s$ should be equal to unity, but we let it as a parameter to be fitted to the simulation results (see below).

The active flow induced by the bacterial suspension $\mathbf{\tilde u}(\tilde t)$ is also modeled as a OUP with
\begin{align}
\langle \tilde u_i (\tilde t) \rangle &= 0,\\
\langle \tilde u_i (\tilde t) \tilde u_i (\tilde t') \rangle &= \tilde u_\text{bh}^2 e^{- |\tilde t - \tilde t'| / \tilde \tau_\text{bh}},
\end{align}
where we have defined $u_\text{bh}$ and $\tau_\text{bh}$ as the velocity and memory time of the bath in the harmonic model, and the tilded variables as their respective dimensionless counterparts.

The solution of Eq.~\eqref{eq.spring_model} is
\begin{equation} \label{eq.solhamornic}
\tilde \rho_i (\tilde t) = \int_{- \infty}^{\tilde t}e^{- (\tilde t - \tilde s) / \tilde \tau_\text{sh}} \tilde u_i (\tilde s)\, d \tilde s,
\end{equation}
where we use that, as the particle is confined, the integral can be done from minus infinity without divergences, avoiding then the use of initial conditions. Using the statistical properties of $\mathbf{\tilde u}$, the mean square displacement is found to be~\cite{szamelSelfpropelledParticleExternal2014,maggiGeneralizedEnergyEquipartition2014}
\begin{align}
\langle \Delta \tilde \rho_i^2 (\tilde t) \rangle &= \frac{2 \tilde u_\text{bh}^2 \tilde \tau_\text{sh}^2 \tilde \tau_\text{bh}}{\tilde \tau_\text{sh}^2 - \tilde \tau_\text{bh}^2} \times \notag \\
& \left[\tilde \tau_\text{sh}\left(1 - e^{- \tilde t / \tilde \tau_\text{sh}} \right) - \tilde \tau_\text{bh} \left( 1 - e^{- \tilde t / \tilde \tau_\text{bh}} \right) \right]. \label{eq.MSD_spring}
\end{align}
In the short time limit the MSD is ballistic, $\lim \limits_{\tilde t \to 0} \langle \Delta \tilde \rho_i^2 (\tilde t) \rangle = {\tilde u_\text{bh}^2 \tilde \tau_\text{sh} \tilde t^2} / (\tilde \tau_\text{sh} + \tilde \tau_\text{bh} )$, and for long times it saturates to $\lim \limits_{\tilde t \to \infty} \langle \Delta \tilde \rho_i^2 (\tilde t) \rangle ={2 \tilde u_\text{bh}^2 \tilde \tau_\text{bh}\tilde \tau_\text{sh}^2}/(\tilde \tau_\text{sh} + \tilde \tau_\text{bh})$.

Similarly, the time-delayed velocity autocorrelation function is
\begin{align}
C_{\tilde vi}(\tilde t, \tilde t')&= \langle \dot{\tilde \rho}_i(t) \dot{\tilde \rho}_i (\tilde t') \rangle,\nonumber \\
&= \left \langle \left( \tilde u_i (\tilde t) - \frac{\tilde \rho_i (\tilde t)}{\tilde \tau_\text{sh}} \right)\left(\tilde u_i(\tilde t') - \frac{\tilde \rho_i (\tilde t')}{\tilde \tau_\text{sh}} \right) \right \rangle,\label{eq.VACF1}\\
&=\frac{\tilde u_\text{bh}^2 \tilde \tau_\text{sh}}{\tilde \tau_\text{sh}^2 - \tilde \tau_\text{bh}^2} \left(\tilde \tau_\text{sh} e^{- |\tilde t - \tilde t'|/\tilde \tau_\text{bh}} - \tilde \tau_\text{bh} e^{- |\tilde t - \tilde t'|/\tilde \tau_\text{sh}} \right). \label{eq.VACF}
\end{align}
This two-exponential function captures the long time decay and the presence of a single negative minimum observed in the simulations (Fig.~\ref{fig.MSD_VACF_PACF}-bottom). It also satisfies the zero time integral demanded by the confinement. Evaluating it at equal times gives the mean square velocity of the particle
\begin{equation}
\label{eq.vacf_tt}
\langle \tilde v_i^2\rangle= C_{\tilde vi}(\tilde t, \tilde t) = \frac{\tilde u_\text{bh}^2 \tilde \tau_\text{sh}}{\tilde \tau_\text{sh} + \tilde \tau_\text{bh}},
\end{equation}
which is consistent with the velocity of the ballistic regime found from the MSD. Interestingly, the particle velocity is different from that of the bath, meaning that it does not thermalize except when the memory of the bath is negligible. Indeed, thermalization for this limiting case is expected as it merely corresponds to a bath in thermal equilibrium. In all other cases, the particle moves at smaller velocities than the bath as a result of the action of the restoring force. In fact, for a bath that pushes persistently ($\tilde\tau_\text{bh} > 0$), it is more probable that the particle position in the trap has the same sign as the instantaneous forcing bath velocity. Mathematically, this can verified using Eq.~\eqref{eq.solhamornic} to evaluate the correlation $\langle \tilde\rho_i(\tilde t)\tilde u_i(\tilde t)\rangle= \tilde u_\text{bh}^2 \tilde \tau_\text{sh}\tilde \tau_\text{bh}/(\tilde \tau_\text{sh} + \tilde \tau_\text{bh})$, which is indeed positive whenever the bath memory is finite. Then, by writing Eq.~\eqref{eq.spring_model} as $\tilde v_i = \tilde u_i - \tilde \rho_i / \tilde{\tau}_\text{sh}$, it is immediate to see that on average the particle speed will be smaller than that of the bath.

\begin{figure*}[ht]

\includegraphics[width=0.32\textwidth]{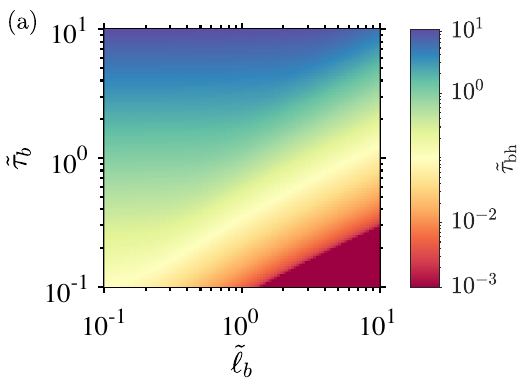}
\includegraphics[width=0.32\textwidth]{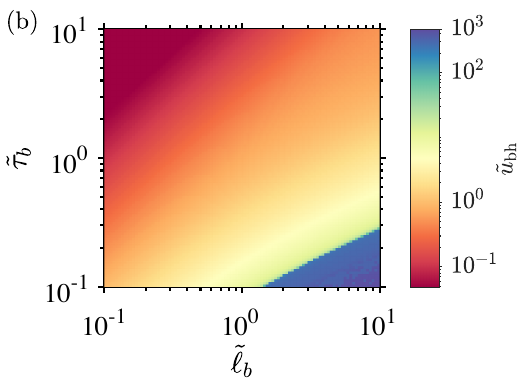}\\
\includegraphics[width=0.32\textwidth]{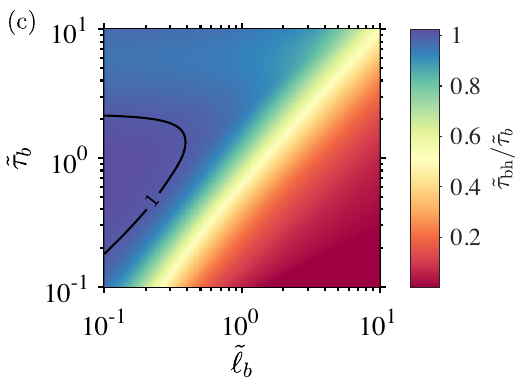}
\includegraphics[width=0.32\textwidth]{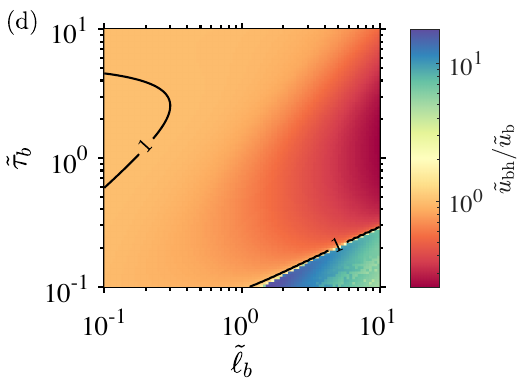}
\includegraphics[width=0.32\textwidth]{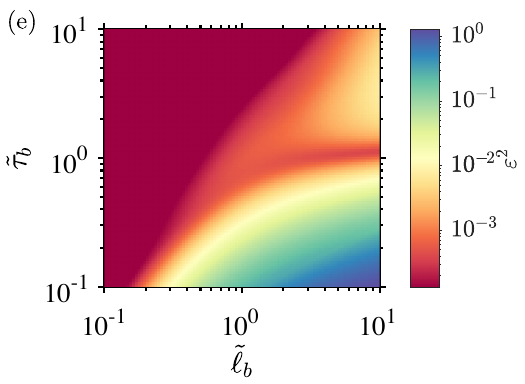}
\caption{Fitted parameters for the harmonic model in the $(\tilde \ell_b$, $\tilde \tau_b)$ space, when the time scale $\tilde\tau_\text{sh}$ is fixed to one (protocol 1). (a)-(b): Fitted values for the harmonic model parameters $\tilde\tau_\text{bh}$ and $\tilde u_\text{bh}$. (c)-(d): Fitted values for the harmonic model parameters normalized to the values in the simulations $\tilde\tau_\text{bh}/\tilde \tau_b$ and $\tilde u_\text{bh}/\tilde{u}_b$. The level curves at 1, that is when the fitted parameters match the imposed ones, are shown in black. (e): mean square error of the fit.}
\label{fig.fitparameters-fix}
\end{figure*}

The harmonic model yields closed-form analytical functions for the MSD and the velocity correlation function that can be used to extract the parameters of the active bath. However, as the two descriptions are not equivalent (one being three dimensional in a confined geometry and the other two dimensional with a harmonic force), the mapping is not expected to be perfect. Moreover, we note that, apart from a trivial rescaling in $\tilde u_\text{bh}$, the two time scales $\tilde \tau_\text{sh}$ and $\tilde \tau_\text{bh}$ play a symmetric role in the mean square displacement ~\eqref{eq.MSD_spring} and the velocity correlation function~\eqref{eq.VACF} expressions. Consequently, by observing the temporal evolution of the tracer and fitting it to these expressions, it would be impossible, in principle, to determine which of the two exponential time scales corresponds to $\tilde \tau_\text{sh}$ or $\tilde \tau_\text{bh}$, unless additional information is available. Experimentally, this can be a subtle point that should be considered with care. In the case we are considering here, we break the symmetry between the two time scales by using that, by the dimensional analysis, $\tilde \tau_\text{sh}$ should be one or close to it.

We fit the harmonic model [Eq.~\eqref{eq.MSD_spring}] to the MSD curves for the planar directions obtained from the simulations of the three-dimensional confined system. For that, we minimize the mean square error defined as
\begin{align}
\varepsilon^2 = \frac{1}{n} \sum_{i=1}^n \left[\log_{10} \text{MSD}_\text{sim}(t_i) -\log_{10} \text{MSD}_\text{model}(t_i) \right]^2,
\end{align}
where, to give equal weights to the different part, the $n$ evaluation times $t_i$ are sampled uniformly distributed in a logarithmic scale (see Ref.~\cite{bailey2022fitting} for a discussion on possible issues when fitting MSD curves).

\begin{figure*}[htb]

\includegraphics[width=0.32\textwidth]{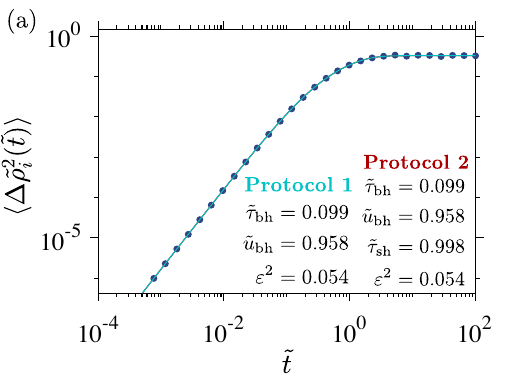}
\includegraphics[width=0.32\textwidth]{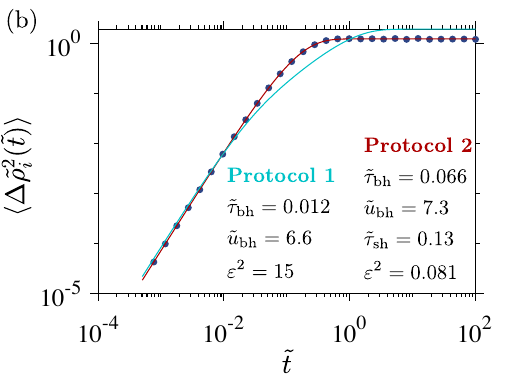}
\includegraphics[width=0.32\textwidth]{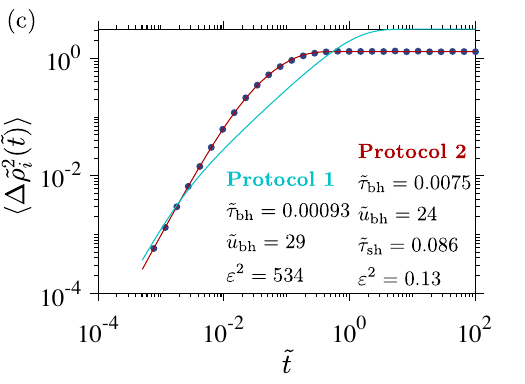}

\caption{Examples of the fit procedure of the MSD using both protocols. The simulation results are shown as back dots (sampled logarithmically in time) and best fit of the model \eqref{eq.MSD_spring} using the first protocol (fixing $\tilde \tau_\text{sh}$ to one) with a cyan line and the second protocol (leaving both times as fitted parameters) with a red line. From left to right, $\tilde \ell_b=0.1,$ 1.0, and 10, while $\tilde \tau_b$ is fixed to 0.1. Insets: fitted parameters and the mean square errors.}
\label{fig.fitprocedure}
\end{figure*}

As a first protocol, we use $\tilde u_\text{bh}$ and $\tilde \tau_\text{bh}$ as fitting parameters, while $\tilde \tau_\text{sh}$ is fixed to one. The results are shown in Fig.~\ref{fig.fitparameters-fix} as a function of the system parameters $\tilde \ell_b$ and $\tilde \tau_b$.
When the harmonic model is a good approximation (region above the solid line in  Fig.~\ref{fig.variance-aprox}-b), one expects to find that $\tilde u_\text{bh} = \tilde u_b$ and $\tilde \tau_\text{bh} = \tilde \tau_b$. Indeed the results in Fig.~\ref{fig.fitparameters-fix} confirm this, showing that the harmonic model is a quantitatively good approximation for the system when $\tilde \tau_b$ is large and $\tilde \ell_b$ is small. Moving away from the parameter region where the position variance is small, the fitted parameters start to deviate strongly from  their expected values. More importantly, the quality of the fit degrades considerably, with large mean square errors. This is manifest when comparing the best fitted MSD to the simulation one, shown in Fig.~\ref{fig.fitprocedure}. The imposed condition that $\tilde \tau_\text{sh} = 1$ makes that for $\tilde \ell_b \gtrsim 1.0$, it is impossible to obtain a good fit, and the obtained parameters lack of any meaning.

To overcome the difficulties stated above, we use a second protocol to obtain the model parameters. Here, $\tilde u_\text{bh}$, $\tilde \tau_\text{bh}$, and $\tilde \tau_\text{sh}$ are all fitting parameters. To break the symmetry between the two time scales, we choose $\tilde \tau_\text{sh}$ as the one that is closest to one. That is, if $|\tilde \tau_\text{sh} - 1| \leq | \tilde \tau_\text{bh}-1|$, the fitted parameters are left unchanged, but on the opposite case, the two times are exchanged and the bath velocity is rescaled as $\tilde{u}_\text{bh} \to \tilde{u}_\text{bh} \sqrt{\tilde{\tau}_\text{sh} / \tilde{\tau}_\text{bh}}$, as shown by Eq.~\eqref{eq.MSD_spring}.

\begin{figure*}[t]

\includegraphics[width=0.32\textwidth]{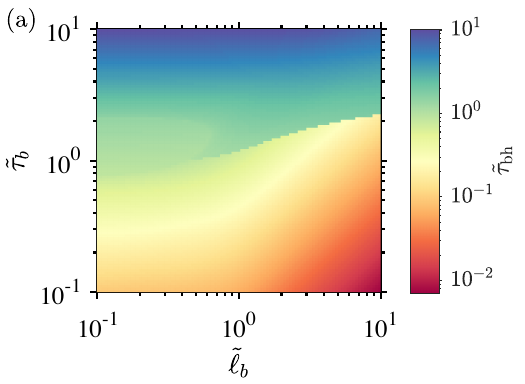}
\includegraphics[width=0.32\textwidth]{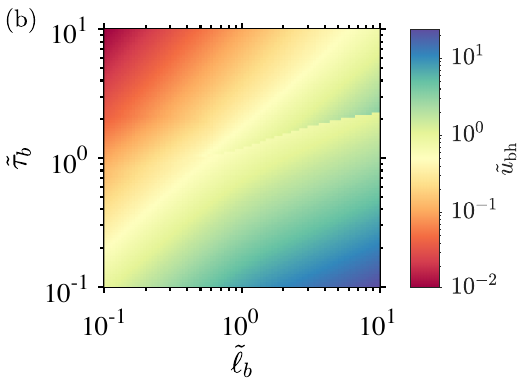}
\includegraphics[width=0.32\textwidth]{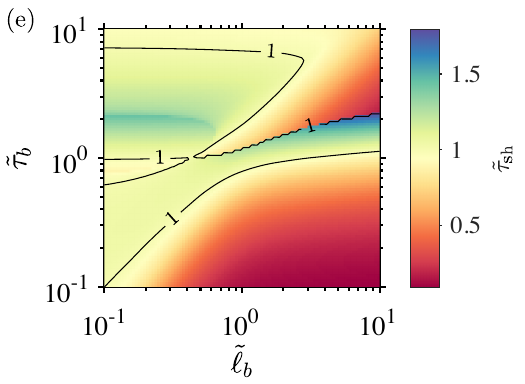}

\includegraphics[width=0.32\textwidth]{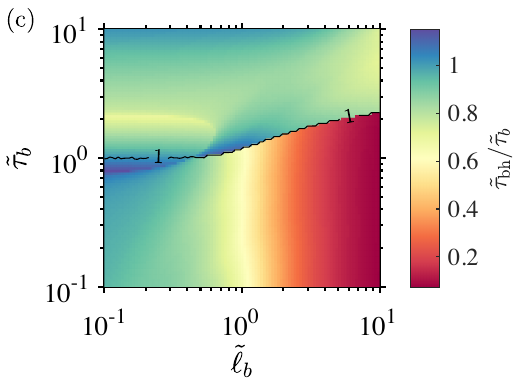}
\includegraphics[width=0.32\textwidth]{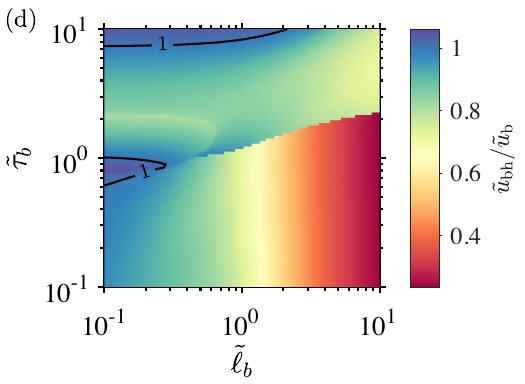}
\includegraphics[width=0.32\textwidth]{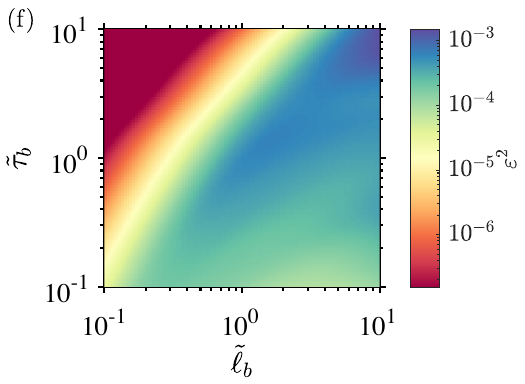}

\caption{Fitted parameters for the harmonic model in the $(\tilde \ell_b$, $\tilde \tau_b)$ space, will all model parameters free and choosing $\tilde\tau_\text{sh}$ to be the closest to one (protocol 2). (a)-(c): harmonic model parameters $\tilde\tau_\text{bh}$, $\tilde u_\text{bh}$, and $\tilde\tau_\text{sh}$. (d)-(e):  harmonic model parameters normalized to their expected values $\tilde\tau_\text{bh}/\tilde \tau_b$ and $\tilde u_\text{bh}/\tilde{u}_b$. (f):  mean square error of the fit.}
\label{fig.fitparameters-flipped}
\end{figure*}

The resulting fitted parameters as well as the mean square error are presented in Fig.~\ref{fig.fitparameters-flipped}. Noticeably, as the condition to exchange the two fitted times is non-analytic, the fitted parameters are also non-analytic in the parameter space and discontinuities appear where both fitted times are close to unity. We first note that the mean square error of the fits is dramatically reduced and that in all the explored parameter space it takes values smaller than $10^{-3}$. This is corroborated by observing the quality of the resulting fits in Fig.~\ref{fig.fitprocedure}, where the simple expression of the harmonic model~\eqref{eq.MSD_spring} is almost indistinguishable with the the simulated MSD for the three dimensional system. Despite this, the fitted parameters do not match those used in the simulations in all the parameter space. However, the region where the agreement is good and the fitting procedure allows to extract the system parameters is much larger when compared to the outcomes of the first protocol. Also, $\tilde \tau_\text{sh}$, which was let as a free parameter, still took values close to one. In the region $\tilde \ell_b \gtrsim 1$ and $\tilde \tau_b \lesssim 1$, both bath parameters $\tilde \tau_\text{bh}$ and $\tilde u_\text{bh}$, as well as the sedimentation time $\tilde \tau_\text{sh}$, are underestimated. Indeed, in this region the harmonic approximation is poor, with the particle largely  departing from the equilibrium position (Fig.~\ref{fig.variance-aprox}).

In summary, although the full dynamics is very different of a simple harmonic model, the MSD is extremely well captured by the linear dynamics represented by the  model, but the parameters do not exactly match.

\section{Discussion}

In this work, we have shown that the parameters characterizing the activity  of a colored noise bath can be extracted from the dynamics of a passive probe particle immersed in the bath enclosed by a spherical confinement.
Notably, despite the motion of the tracer being three-dimensional, it is possible to extract the bath parameters using only the information of the two-dimensional motion.
 For that, we fitted the planar MSD to a curve obtained from the analytical solution of a particle trapped in a harmonic potential that deviates from the equilibrium position when driven by the active bath. Although such harmonic trap approximation is expected to hold only when the probe particle remains near the equilibrium position, we find that this strategy yields reasonable results even when the particle explores a rather large region of the sphere, up to variances $\langle \rho^2 \rangle \lesssim 0.5R^2$. In principle, the parameters of the active fluid could be obtained similarly from a fit of the velocity correlation function to the analytical from stemming from the harmonic model, Eq.~\eqref{eq.VACF}. However,  the particle velocities are not directly accessible but are rather obtained numerically by differentiating the sampled particle positions, and hence  would lead in practice to noisier results.

Importantly, in the harmonic model, the memory time of the flow and the relaxation time associated with the confinement play a symmetric role in the expressions for the MSD and velocity correlation function. This symmetry impedes in principle to distinguish between these two time scales by only observing the dynamics of the particle. Additional information is needed to break this degeneracy, which in the case of the particle confined inside the sphere was provided by our knowledge of the sedimentation time.

The model we consider has a constant sedimentation speed, independent on the particle position, meaning that we neglect hydrodynamic lubrication effects. This approximation is consistent with considering that the induced flow $\mathbf{u}$ is independent of particle position, which allowed us to model $\mathbf{u}(t)$ as a stationary autonomous stochastic process. For all simulation conditions, with the sole exception of very large bath velocities $\tilde u_b$, the simulations show that the probe particle remains almost all the time in contact with the surface. This implies that the friction coefficient of the inner particle may not be that of a particle in the bulk, but eventually modified by hydrodynamic lubrication. This value depends logarithmically on the gap distance $R_o - R_i$~\cite{Dunstan2012,aponte2016simulation} and therefore may change on time. Hence, in our model, $\gamma$ should be understood as the friction coefficient averaged over the possible gap values. Interestingly, the second fitting protocol we present, leaving the sedimentation time as a free parameter, allows to obtain the effective friction coefficient. However, to apply the protocol, some prior estimation of its value should be available to break the symmetry between the two times obtained from the fit.

\section*{Acknowledgments}
This research was supported by the Millennium Science Initiative Program NCN19\_170 and Fondecyt Grants No.\ 1220536 (R.S.) and 1210634 (M.L.C.), all from ANID, Chile, and the Franco-Chilean Ecos-Sud Collaborative Program 210012. C.V.C acknowledges funding from ANID Beca de Doctorado Nacional 21201766.
E.C. thanks the support of the Action program CNRS-MITI-2022:Auto-Organisation, the ANR-22-CE30 grant ``Push-pull'' and the Institut Universitaire de France (IUF).

´

%

\end{document}